# Symmetry, degeneracy and optical confinement of modes in coupled microdisk resonators and photonic crystal cavities

Svetlana V. Boriskina, *Member*, IEEE

**Journal-ref:** submitted to *IEEE J. Quantum Electron.*, Sept 2005.

*Abstract*— **Fast and highly accurate method based on the Muller contour integral equations and a trigonometric-trigonometric Galerkin discretization technique is presented for spectral design and fine-tuning of clusters of whispering-gallery (WG) mode microdisk resonators and photonic crystal (PC) defect cavities. It is shown that degeneracy and near-degeneracy of natural modes as well as modes optical confinement in coupled resonator clusters are significantly influenced by the symmetry of the structure. Photonic molecules with greatly enhanced quality factors are designed by exploiting the symmetry of the molecular geometry and tuning the inter-cavity coupling distances. Potential applications of the spectrally engineered coupled-resonator structures to the low-threshold microdisk lasers, optical biosensors, and random powder lasers are discussed.**

## I. Introduction

HIGH-Q semiconductor microdisks and photonic crystal defect cavities are essential for the development of the next generation of optoelectronic components such as light emitting diodes, low threshold microlasers, narrow-linewidth wavelength selective filters, biochemical sensors, etc [1, 2]. Arranging several electromagnetically coupled microcavities into so-called 'photonic molecules' offers a possibility of achieving new functionalities of the devices without compromising the high Q-factors of individual cavities [3].

Advanced lithographic techniques now enable robust fabrication of semiconductor photonic molecules of various shapes and sizes [3, 4]. Complex 2-D and 3-D coupled-resonator structures can also be formed by micromanipulation or self-assembly of polystyrene microspheres [5, 6] and by linking together polymer-blend microparticles generated from microdroplets of dilute solution [7]. Among various practical applications of coupled-microcavity structures are low-threshold microlasers [4, 8, 9], mode switching [4], high-order optical waveguide filters [10, 11], and optical power transfer trough coupled-resonator optical waveguides (CROWs) [5, 7, 9]. Furthermore, coupled-microcavity models can be useful for interpreting narrow peaks observed in experimental spectra of random powder lasers when a powder particle size is comparable to or larger than the emission wavelength [12].

However, micro- to nano-scale sizes and high index contrasts make accurate and efficient modeling of coupled-resonator structures computationally challenging. Simple and flexible finite-difference time-domain method, while being a main workhorse in simulation of optical microcavities [4, 13], may suffer from staircasing errors or become prohibitively computationally expensive for complex cavity geometries. Other popular numerical tools such as the coupled mode theory and the tight binding approximation are able to provide fast initial designs of coupled-microcavity structures, but have uncontrollable accuracy.

Simulation tools based on a rigorous integral-equation formulation used in this paper, which take into account all the mutual electromagnetic interactions in the system, have already been proven to yield accurate solutions to a variety of design problems in photonics [14]. They provide not only faster results but also insights into physics of optical resonator clusters and general design rules to improve their performance for a variety of existing and emerging applications.

## II. Methodology

We consider an eigenvalue problem for a set of closely located side-coupled microdisk resonators as illustrated in Fig. 1. Due to strong vertical confinement in thin high-index-contrast microdisks, it is possible to separate the modal field solutions into two polarizations: TE and TM, which have the electric (magnetic) field polarized in the plane of the microdisk. All the electromagnetic field components for the TM (TE) polarization can be expressed through a single $z$-component of either electrical or magnetic field, which satisfy the 2-D Helmholtz equation, field continuity conditions at all the material interfaces, and a Reichardt's condition at infinity [15]. The effective values of microdisk refractive indices used in the following 2-D analysis are calculated with the effective index method, which assumes that the microdisk modes have the same spatial dependence in the vertical direction as guided modes of an equivalent slab waveguide [14].

S. V. Boriskina was with the George Green Institute for Electromagnetics Research, University of Nottingham, Nottingham. She now is with the School of Radiophysics, V. Karazin Kharkov National University, Kharkov, Ukraine (e-mail: SBoriskina@gmail.com).

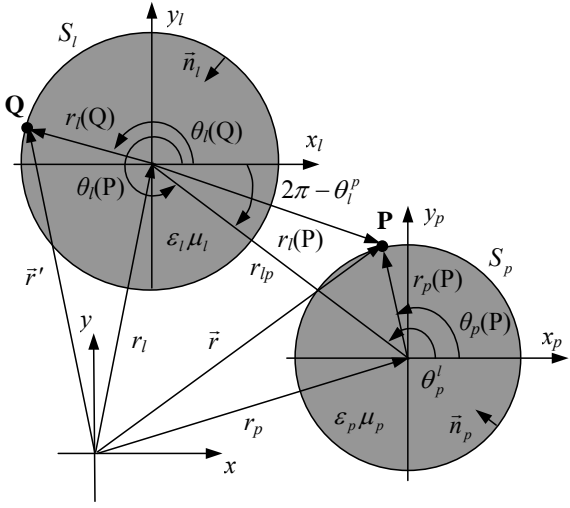

Fig. 1. Schematic of two closely located microcavities together with the global and local coordinate systems used in the analysis.

Applying the Green's formula to the fields and the Green's functions in the regions inside and outside all the microcavities and taking into account the conditions at the cavity boundaries, we formulate the problem in terms of the Muller boundary integral equations (MBIEs):

$$U_p(\vec{r}) = U^{inc}(\vec{r})$$
$$+ \sum_{l=1}^{L} \int_{S_l} \left[ U_l(\vec{r}') \frac{\partial}{\partial n'} \left( G_l(\vec{r},\vec{r}') - G_e(\vec{r},\vec{r}') \right) \right.$$
$$\left. - V_l(\vec{r}') \left( G_l(\vec{r},\vec{r}') - \frac{\alpha_e}{\alpha_l} G_e(\vec{r},\vec{r}') \right) \right] ds'_l$$

$$\frac{\alpha_e + \alpha_p}{2\alpha_p} V_p(\vec{r}) = V^{inc}(\vec{r})$$
$$+ \sum_{l=1}^{L} \int_{S_l} \left[ U_l(\vec{r}') \frac{\partial^2}{\partial n \partial n'} \left( G_l(\vec{r},\vec{r}') - G_e(\vec{r},\vec{r}') \right) \right.$$
$$\left. - V_l(\vec{r}') \frac{\partial}{\partial n} \left( G_l(\vec{r},\vec{r}') - \frac{\alpha_e}{\alpha_l} G_e(\vec{r},\vec{r}') \right) \right] ds'_l$$

(1)

where $p = 1...L$ ($L$ being a total number of cavities) and

$$U_l^{TE}(\vec{r}) = H_z^l(\vec{r}), \quad V_l^{TE}(\vec{r}) = \frac{\partial H_z^l(\vec{r})}{\partial n}, \quad \alpha_l = \varepsilon_l, \quad \alpha_e = \varepsilon_e,$$
$$U_l^{TM}(\vec{r}) = E_z^l(\vec{r}), \quad V_l^{TM}(\vec{r}) = \frac{\partial E_z^l(\vec{r})}{\partial n}, \quad \alpha_l = \mu_l, \quad \alpha_e = \mu_e.$$

Equations (1) are the Fredholm second-kind integral equations with smooth or integrable kernels and are free of spurious solutions [14, 16]. Thus, their discretization yields well-conditioned matrices. For the case of a circular 2-D microcavity of radius $a_p$ with the source (Q) and observation (P) points belonging to the same $p$-th cavity ($Q \in S_p$, $P \in S_p$), the kernel Green's functions can be written in the coordinate system associated with the $p$-th cavity as follows:

$$G_p(\vec{r},\vec{r}') = \frac{i}{4} \sum_{(m)} J_m(k_p r_p(P)) H_m^{(1)}(k_p r_p(Q)) e^{-im\theta_p(Q)} e^{im\theta_p(P)}$$
$$G_e(\vec{r},\vec{r}') = \frac{i}{4} \sum_{(m)} J_m(k_e r_p(Q)) H_m^{(1)}(k_e r_p(P)) e^{-im\theta_p(Q)} e^{im\theta_p(P)}$$

(2)

Similar expressions for the Green's functions are valid if the source and the observation points belong to different cavities ($Q \in S_l$, $P \in S_p$), however in this case the Graf's formulas should be used to transform the expressions for $l = 1...L$ to a unique coordinate system associated with the $p$-th cavity [17]:

$$H_m^{(1)}(kr_l(P))e^{im\theta_l(P)} = \begin{cases} \sum_{(n)} H_{n-m}^{(1)}(kr_{pl}) J_n(kr_p(P)) e^{i(m-n)\theta_p^l} e^{in\theta_p(P)} \\ \text{for } r_p(P) < r_{pl} \\ \sum_{(n)} J_{n-m}(kr_{pl}) H_n^{(1)}(kr_p(P)) e^{i(m-n)\theta_p^l} e^{in\theta_p(P)} \\ \text{for } r_p(P) > r_{pl} \end{cases}$$

(3)

Now we substitute the above expressions into the MBIEs (1), and expand the unknown functions in terms of the Fourier series with angular exponents as global basis functions. Testing against the same set of global functions yields the final matrix equation that needs to be solved to determine the Fourier coefficients $u_m^p$ and $v_m^p$ (see [14] for more detail):

$$a_m^p u_m^p + b_m^p v_m^p + \sum_{l \neq p} \left\{ \sum_{(n)} u_n^l A_{mn} + \sum_{(n)} v_n^l B_{mn} \right\} = 0$$
$$c_m^p u_m^p + d_m^p v_m^p + \sum_{l \neq p} \left\{ \sum_{(n)} u_n^l C_{mn} + \sum_{(n)} v_n^l D_{mn} \right\} = 0$$

(4)

where

$$a_m^p = \sqrt{\varepsilon_p} J_m'(k_p a_p) H_m^{(1)}(k_e a_p)$$
$$- \sqrt{\varepsilon_e} J_m'(k_e a_p) H_m^{(1)}(k_e a_p) + \frac{4}{i\pi k a_p}$$

$$b_m^p = J_m(k_p a_p) H_m^{(1)}(k_e a_p) - \frac{\alpha_e}{\alpha_p} J_m(k_e a_p) H_m^{(1)}(k_e a_p)$$

$$c_m^p = \varepsilon_e J_m'(k_e a_p) H_m^{(1)'}(k_e a_p) - \varepsilon_p J_m'(k_p a_p) H_m^{(1)'}(k_p a_p)$$

$$d_m^p = \frac{\alpha_e}{\alpha_p} \sqrt{\varepsilon_e} J_m(k_e a_p) H_m^{(1)'}(k_e a_p)$$
$$- \sqrt{\varepsilon_p} J_m'(k_p a_p) H_m^{(1)}(k_p a_p) + \frac{2(\alpha_p + \alpha_e)}{i\pi \alpha_p k a_p}$$

and

$$A_{mn} = \left( \sqrt{\varepsilon_l} J_n'(k_l a_l) J_m(k_l a_p) H_{m-n}^{(1)}(k_l r_{pl}) \right.$$
$$\left. - \sqrt{\varepsilon_e} J_n'(k_e a_l) J_m(k_e a_p) H_{m-n}^{(1)}(k_e r_{pl}) \right) e^{i(n-m)\theta_p^l}$$

$$B_{mn} = \Big(J_n(k_l a_l)J_m(k_l a_p)H^{(1)}_{m-n}(k_l r_{pl})$$
$$- \frac{\alpha_e}{\alpha_l} J_n(k_e a_l)J_m(k_e a_p)H^{(1)}_{m-n}(k_e r_{pl})\Big)e^{i(n-m)\theta^l_p}$$

$$C_{mn} = \Big(\varepsilon_e J'_n(k_e a_l)J'_m(k_e a_p)H^{(1)}_{m-n}(k_e r_{pl})$$
$$- \varepsilon_l J'_n(k_l a_l)J'_m(k_l a_p)H^{(1)}_{m-n}(k_l r_{pl})\Big)e^{i(n-m)\theta^l_p}$$

$$D_{mn} = \Big(\frac{\alpha_e}{\alpha_l}\sqrt{\varepsilon_e} J_n(k_e a_l)J'_m(k_e a_p)H^{(1)}_{m-n}(k_e r_{pl})$$
$$- \sqrt{\varepsilon_l} J_n(k_l a_l)J'_m(k_l a_p)H^{(1)}_{m-n}(k_l r_{pl})\Big)e^{i(n-m)\theta^l_p}$$

Homogeneous equation (4) has nonzero solutions only at frequencies where the equation matrix becomes singular. Complex eigenfrequencies of the optical modes supported by a set of coupled resonators are found by searching for the roots of the matrix determinant in the complex frequency plane. Once the eigenfrequencies are found, the modal fields and emission characteristics can be calculated by solving the homogeneous matrix equation (4) at the eigenfrequencies.

### III. ALGORITHM VALIDATION AND ACCURACY

First, we test the developed algorithms against the results of finite-difference time-domain (FDTD) simulations available in the literature. A finite-size photonic crystal cavity composed of 25 GaAs rods in air is chosen as a test structure (Fig. 2). This PC cavity has four TM defect modes in the bandgap (from 0.29 $d/\lambda$ to 0.42 $d/\lambda$): non-degenerate OO and EE quadrupole modes (Fig. 2 a, b), a non-degenerate second-order monopole mode (Fig. 2 c), and a double-degenerate (OE and EO) hexapole mode (Fig. 2 d). Here, the modes are classified in accordance with their field symmetry relative to the $x$ and $y$ axes (e.g., OE mode is anti-symmetrical with respect to the $x$ axis and symmetrical with respect to the $y$ axis).

TABLE 1. COMPARISON OF MBIES AND FDTD FOR A PC CAVITY

| Mode | Normalized frequency, $d/\lambda$ | | Quality factor, Q | |
|---|---|---|---|---|
| | FDTD [13] | MBIEs | FDTD | MBIEs |
| OO quadrupole | 0.2970 | 0.2967 | 648 | 855 |
| EE quadrupole | 0.3190 | 0.3196 | 276 | 281 |
| Monopole | 0.3345 | 0.3349 | 466 | 488 |
| Hexapole | 0.3916 | 0.3916 | 2936 | 4887 |

The resonant frequencies and Q-factors of the four PC cavity modes computed with MBIEs algorithm and a standard 2-nd order FDTD scheme [13] with Yee's cell of $(25\text{ nm})^2$ are presented in Table 1. Thanks to exponential convergence of the MBIEs algorithm, the guaranteed uniqueness of solution, and the choice of the Green's functions satisfying the radiation condition at infinity, the MBIEs results in Table 1 have been obtained with the machine precision. It can be seen that the values of the resonant frequencies computed with both methods are in good agreement. However, as the Q-factors grow, failing ability of the FDTD algorithm with the same spatial grid to reproduce their values accurately becomes evident. To reach the same accuracy of the FDTD solution for the high-Q hexapole mode as for the lower-Q monopole, the problem matrix size (and thus the computer resources) will have to be increased significantly due to the polynomial convergence rate of the FDTD scheme.

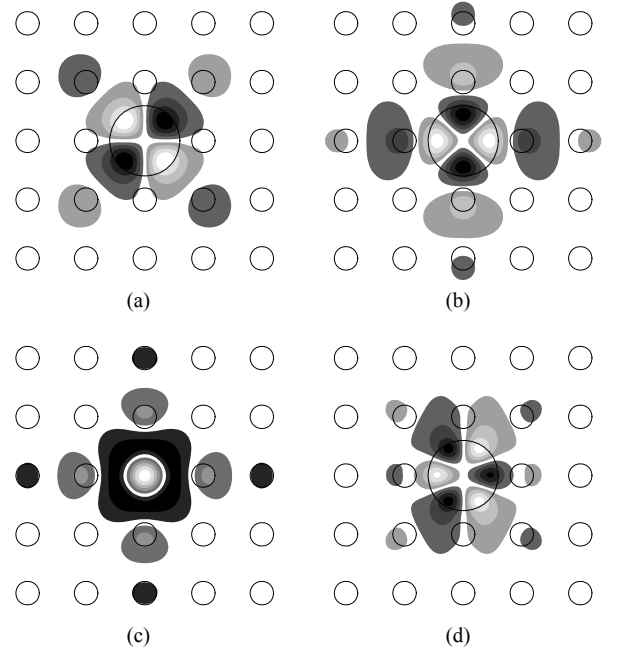

Fig. 2. The electric field profiles ($E_z$) of the (a) OO quadrupole, (b) EE quadrupole, (c) monopole, and (d) EO hexapole modes of a square-lattice photonic crystal cavity composed of GaAs ($\varepsilon$=11.56) rods of 0.2 μm radii and a single defect rod of 0.6 μm radius.

Having tested the algorithm and software, we will now apply it to study optical properties of coupled-microcavity structures of various shapes with the aim to establish design rules to form photonic molecules with improved performance.

### IV. SPECTRAL ENGINEERING OF MICROCAVITY CLUSTERS

#### A. Optical spectra of isolated microcavities

Circular microdisk resonators support dense frequency-dependent TE- and TM- spectra of whispering-gallery (WG) modes confined in the microdisk by the mechanism of the total internal reflection. The TE- (TM-) modes are classified as WGE(H)$_{m,n}$ modes, $m$ being the azimuthal mode number, and $n$ the radial mode number (see, e.g. [1, 2, 14] for detail). Due to the symmetry of the structure, all the WG-modes are double-degenerate, and the number of high- and low-Q WG modes increases with the increase of the cavity optical size.

In photonic crystal defect cavities, the photonic-bandgap effect is used for strong light confinement in the cavity plane, and introduced defects result in appearance of a finite number of localized states in the band gap [1, 2, 13, 18, 19]. Unlike WG-mode microdisk resonators, PC defect cavities can support high-Q non-degenerate modes (Fig. 2 a-d).

#### B. Mode splitting in double-cavity photonic molecules

If two microcavities are electromagnetically coupled (e.g., via a finite-width airgap [4, 8] or a narrow material bridge [3]),

non-degenerate individual cavity modes split into two non-degenerate bonding (symmetrical) and anti-bonding (anti-symmetrical) modes. An example of mode splitting showing near-field portraits of a monopole mode of a single PC cavity as well as bonding and anti-bonding modes of a double-cavity photonic molecule is presented in Fig. 3. The anti-bonding mode is red-shifted and the bonding mode is blue-shifted. These results are in good agreement with the plane-wave scattering spectra for photonic crystals doped with microcavities computed in [18]. Similar theoretical and experimental observations of the mode doublets have been made for non-degenerate zero-order modes of side-coupled circular [19] and square-shaped [3] microdisk resonators.

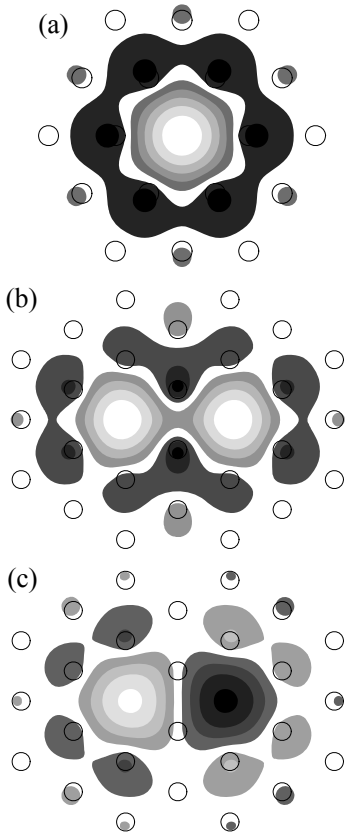

Fig. 3. The electric field profiles ($E_z$) of (a) a non-degenerate monopole mode in a hexagonal photonic crystal defect cavity ($\lambda = 9.054$ μm, Q = 56.8); (b) bonding ($\lambda = 8.837$ μm, Q = 90) and (c) anti-bonding ($\lambda = 9.324$ μm, Q = 67.4) non-degenerate modes of a photonic molecule composed of two coupled hexagonal PC cavities. PC lattice parameters are as follows: $\varepsilon_{rods} = 8.41$, $a = 0.6$ μm, $d = 4.0$ μm.

In turn, degenerate WG-modes of pairs of optical resonators brought together split into two groups of nearly-degenerate photonic molecule modes (see Fig. 4 as an example of double-degenerate $WG_{2,1}$-mode splitting in a double-disk photonic molecule). The number of nearly-degenerate modes in each group depends on the individual cavity mode degeneracy (e.g., two for a circular-disk photonic molecule [20]). The mode with the highest Q-factor is the blue-shifted anti-bonding mode anti-symmetrical with respect to both x- and y-axes. Note that similar degenerate mode splitting occurs when the cavity is located close to a conducting plane, a material interface or a slab waveguide [15]. Geometrical imperfections may further split the modes [21] causing appearance of several closely located peaks in the photonic molecule spectra [22].

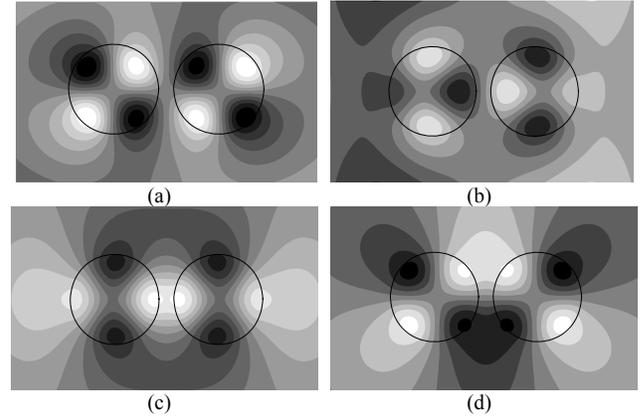

Fig. 4. The electric field profiles ($E_z$) of the four non-degenerate $WGH_{2,1}$-modes in a photonic molecule composed of two identical coupled microdisk resonators with $\varepsilon = 10.24+i10^{-4}$, $a = 0.3$ μm and $d = 0.7$ μm. (a) OO: $\lambda = 1.567$ μm, Q = 38.6; (b) EO: $\lambda = 1.6$ μm, Q = 18.6; (c) EE: $\lambda = 1.661$ μm, Q = 16.6; and (d) OE: $\lambda = 1.711$ μm, Q = 29.6.

The performed careful review of the properties of double-cavity photonic molecules supporting various types of modes gives us a useful insight into the behavior of more complex coupled-microcavity structures. In the following sections, symmetry-enhanced mode optical confinement and degeneracy removal mechanisms are identified through a detailed study of the degenerate mode wavelength splitting and Q-factors change as a function of the photonic molecule geometry and the inter-cavity coupling distances.

### C. Optical spectra of linear photonic molecules

First, linear chains of side-coupled microcavities will be studied and common features of their optical spectra will be identified. Here and in the following sections, the geometrical and material parameters of the resonators are chosen as follows: GaInAsP microdisks have diameters of 1.8 μm and thickness of 200 nm. Only the TE modes with one vertical field variation can be supported in the disk of such thickness, all the TM-polarized modes and higher vertical order TE modes are suppressed. The effective value of the dielectric permittivity ($\varepsilon=6.9169+i10^{-4}$) used in the following 2D analysis corresponds to the propagation constant of the TM-polarized mode in an equivalent slab waveguide at 1.55 μm [14] (spontaneous emission peak of GaInAsP at room temperature is located at 1.55-1.58 μm).

Fig. 5 shows the variations in the modal wavelengths and Q-factors of $WGE_{6,1}$ modes in a linear photonic molecule composed of three identical circular microdisk resonators coupled via airgaps as a function of the inter-cavity coupling distance. One can clearly see (Fig. 5 a) that split modes of different symmetry group into nearly degenerate mode pairs, with bonding modes shifting to longer wavelengths and anti-bonding modes shifting to shorter wavelengths. The number of nearly degenerate mode pairs is equal to the number of microcavities forming a photonic molecule, which is in excellent agreement with experimental data [4]. With the increase of the WG-mode azimuthal order the spacing between nearly degenerate modes decreases, which makes it difficult to distinguish between individual modes in the photonic molecule spectra. However, the modes never become truly degenerate as they have different imaginary parts of the complex natural frequencies (and thus different Q-factors).

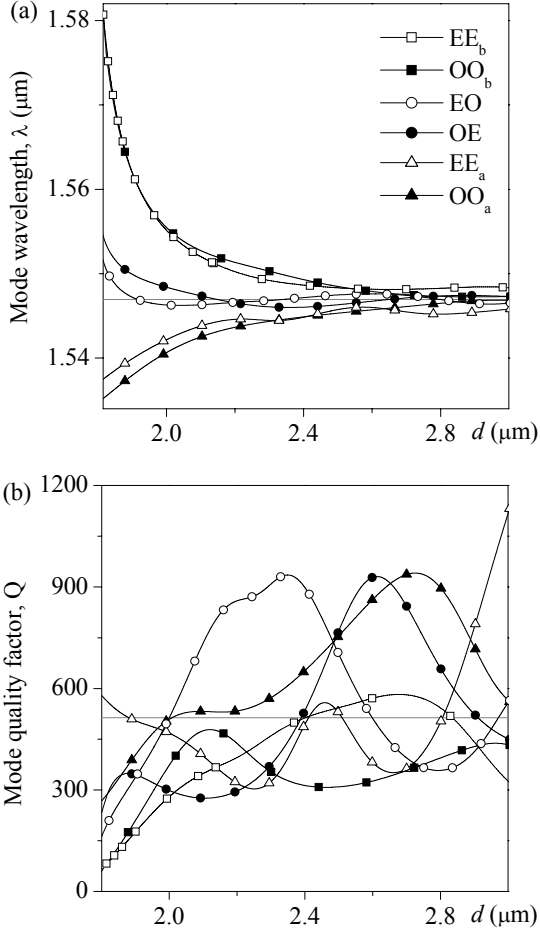

Fig. 5. $WGE_{6,1}$ mode wavelengths splitting (a) and quality factors change (b) in a linear chain of three coupled identical microdisks versus a center-to-center distance between neighboring resonators. Corresponding single-microdisk characteristics are plotted for comparison (straight lines).

The Q-factors of the linear-molecule modes oscillate around the value of the single-cavity Q-factor (Fig. 5b). This phenomenon has been studied theoretically for double-cavity molecules in [20] and [23], though in [23] mode splitting has not been discussed. Thus, suppression or enhancement of modes can be achieved by properly tuning the inter-cavity coupling distance. However, due to the near-degeneracy of the modes, single-mode operation necessary for microlaser applications cannot be achieved in linear photonic molecules.

### D. Symmetry-assisted mode enhancement

In this section, we will study the optical spectra of high-symmetry photonic molecule structures such as triangles, squares and hexagons to reveal the role of symmetry in enhancing the molecule optical performance. In this section, the modes are classified by how they transform under the mirror operation with respect to the symmetry axes of the structure (medians of the triangular molecule, and symmetry lines going through the corners or the sides of the square and hexagonal molecules).

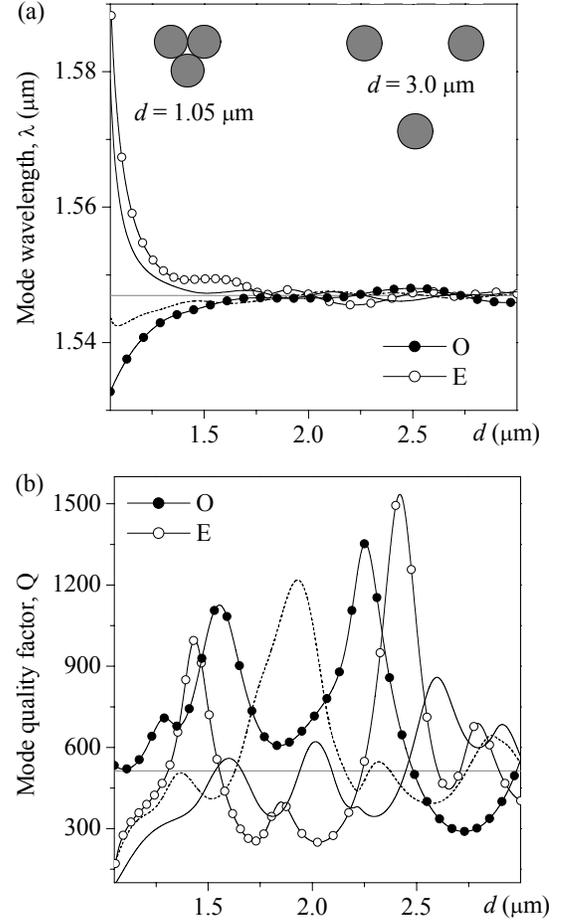

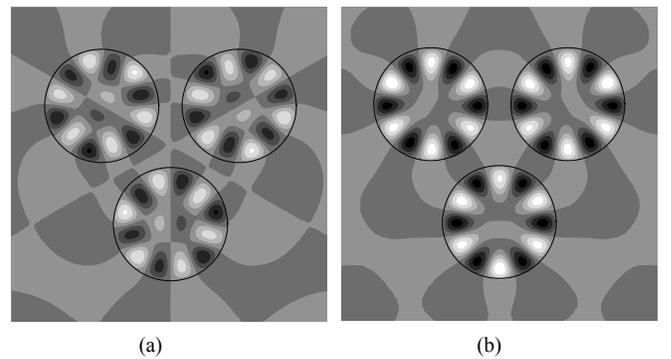

Fig. 6. $WGE_{6,1}$ mode wavelengths splitting (a) and quality factors change (b) in a triangular photonic molecule as a function of a distance from the resonators center to the molecule center. Corresponding single-microdisk characteristics are plotted for comparison (straight lines). The insets show the molecule geometries for $d=1.809$ μm and $d=3$ μm.

Fig. 7. The magnetic field profiles ($H_z$) of the (a) O and (b) E $WGE_{6,1}$-modes of the triangular photonic molecule with $d = 1.25$ μm.

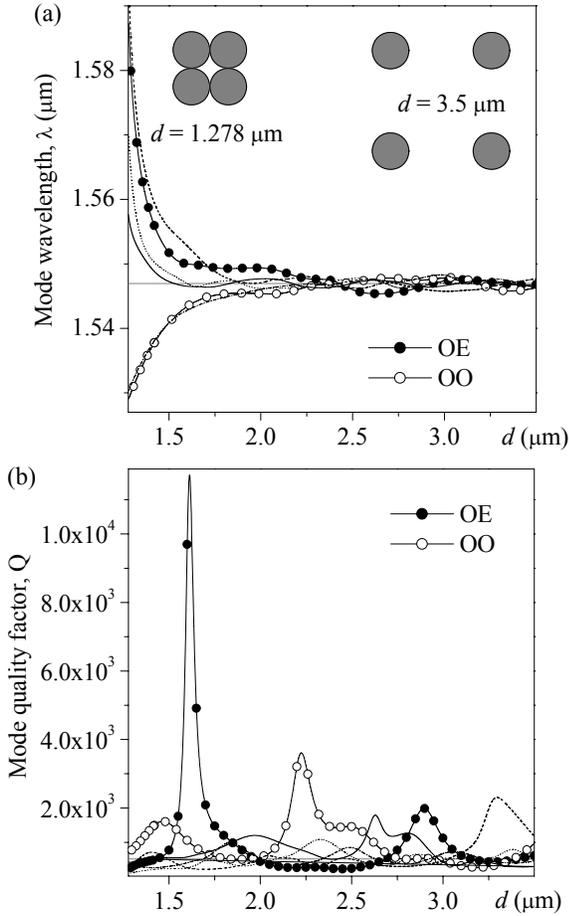

Fig. 8. WGE$_{6,1}$ mode wavelengths splitting (a) and quality factors change (b) in a square photonic molecule as a function of a distance from the resonators center to the molecule center. Corresponding single-microdisk characteristics are plotted for comparison (straight lines). The insets show the molecule geometries for $d$=1.278 μm and $d$=3.5 μm.

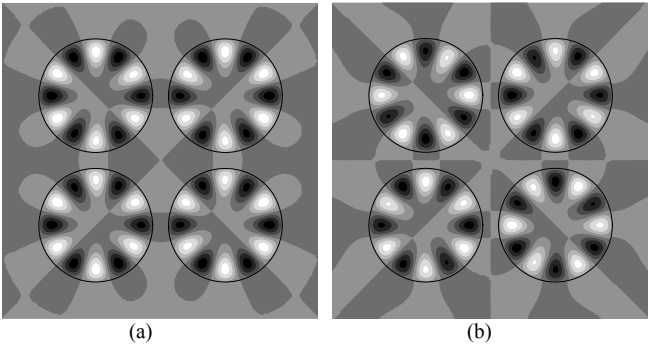

Fig. 9. The magnetic field profiles of the (a) OE and (b) OO WGE$_{6,1}$-modes of the square photonic molecule with $d$ = 1.45 μm.

Fig. 6 shows the modal wavelengths and Q-factors in a photonic molecule composed of three microdisks located at the corners of an equilateral triangle versus the inter-cavity coupling distance. Note that only four modes instead of six are seen in the photonic molecule spectra. Due to the symmetry of the structure, two modes are double-degenerate (shown with solid and dotted lines in Fig. 6). Non-degenerate modes have either odd or even field symmetry along the triangle medians (Fig. 7). An important feature of Fig. 6 b is that the Q-factors of these non-degenerate modes experience the most pronounced increase for certain values of the inter-cavity distance. This effect offers a way to design symmetry-enhanced single-mode high-Q photonic molecule configurations. We find that the distortion of the molecule symmetry results in splitting of the degenerate WG-modes and reduces the effect of the Q-factor enhancement of the non-degenerate ones (which is in agreement with the optical and microwave experimental data for zero- and low-azimuthal-order modes of square- and circular-disk photonic molecules [19, 24]).

Next, photonic molecule structures with higher symmetry will be studied. Fig. 8 presents a square-molecule optical spectra dependence on the change of the coupling distance. Here, four modes are non-degenerate: OO, EE, OE, and EO modes (the first (second) letter denotes the modal symmetry with respect to the diagonals and the $x$- and $y$- axes, respectively). The most significant Q-factor increase (e.g., by a factor of 23 for the OE mode) is observed for the modes with the odd symmetry along the square diagonals (see Fig. 9 for their near-field portraits). It can be clearly seen that higher symmetry of a square molecule yields much higher mode enhancement than a triangular one.

Finally, modal characteristics of a hexagonal-shaped photonic molecule are presented in Fig. 10. Once again, the modes that can be drastically enhanced by carefully tuning the inter-cavity coupling distance are the OO and OE non-degenerate modes with the odd symmetry with respect to the hexagon diagonals. Fig. 11 presents their near-field distributions.

Note that optimally tuned photonic molecule structures have much larger free spectral range than larger-radius individual microdisk resonators with comparable values of Q-factors. Furthermore, as the high-Q modes of the photonic molecules are non-degenerate, such structures are expected to be less sensitive to fabrication imperfections that split WG-modes of single-cavity resonators causing appearance of parasitic peaks in their optical spectra [21]. Arranging circular resonators into photonic molecules also enables higher design flexibility than introducing artificial deformations to the shape of individual resonators [14, 21]. The above observations also reveal the analogy between the WG-modes in symmetrical photonic molecules and the non-degenerate high-Q modes in square and triangular microcavities (e.g. recently observed WG-like modes in square resonators) that have odd symmetry around the cavity diagonals [25, 26].

*E. Applications of spectrally-designed photonic molecules*

The identified properties of the optical spectra of high-symmetry photonic molecules can be advantageous for various applications not only in the fields of optoelectronics and laser physics, but also for micro- and millimeter-wave technology and acoustics. For example, the observed mechanism of the significant mode Q-factor enhancement accompanied by a mode polarization degeneracy removal paves a way for designing and fabricating single-mode photonic-molecule lasers with large spontaneous emission factors and low thresholds. Furthermore, as the mode wavelength is very sensitive to the inter-cavity distances, the molecule geometry can be tuned to spectrally align the mode wavelength with the semiconductor material gain peak thus increasing the fraction of spontaneous emission into this mode.

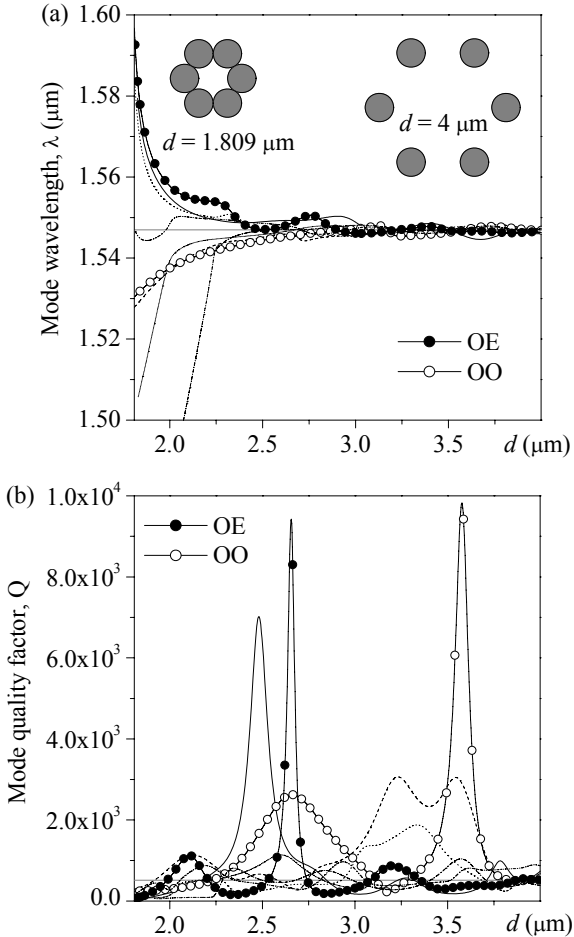

Fig. 10. $WGE_{6,1}$ mode wavelengths splitting (a) and quality factors change (b) in a hexagonal photonic molecule as a function of a distance from the resonators center to the molecule center. Corresponding single-microdisk characteristics are plotted for comparison (straight lines). The insets show the molecule geometries for $d=1.809$ μm and $d=4.0$ μm.

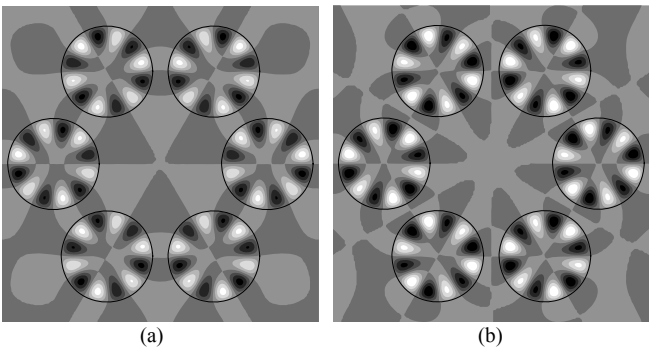

Fig. 11. The magnetic field profiles of the (a) OE and (b) OO $WGE_{6,1}$-modes of the hexagonal photonic molecule with $d = 2.1$ μm.

Due to this high sensitivity of the modal frequencies and Q-factors to the change of the molecule structural parameters, their application as biochemical sensors is expected to offer significantly enhanced sensitivity compared to single-resonator sensors. Such sensors often exploit a detection mechanism based on measuring the resonant frequency shift caused by the presence of a biochemical agent through the interaction of the evanescent field of the WG-mode outside the resonator with the analyte. As the photonic molecule modes are collective multi-cavity resonances, their frequencies and Q-factors should be more sensitive to the change of the refractive index in the inter-cavity region than the corresponding characteristics of a single cavity.

The summarized main features of the coupled-microcavity structures and established design rules to tailor their spectral characteristics can be directly applied to simulate the high-Q localized modes of powder lasers [12] and to design photonic molecules composed of microspheres [5-7, 27] as well as to the fields of microwave engineering [28] and acoustics [29].

V. CONCLUSIONS

In this paper, a systematic theoretical study of the properties of whispering gallery modes in coupled-microcavity clusters of various shapes has been reported. An efficient technique based on the MBIEs method has been developed and used to simulate and fine-tune the clusters optical characteristics as well as to help in interpreting their experimental spectra. The MBIEs method, applied here to study WG-mode circular-microcavity clusters that are essential for a variety of applications, is a powerful general technique enabling efficient modeling of arbitrary-shape resonators [14, 15, 21, 25]. The algorithms, numerical results, and conclusions have been validated by comparing them against the available experimental data and FDTD computations. Several coupled-microcavity structures with symmetry-enhanced Q-factors of non-degenerate modes have been proposed and studied in detail. Among possible applications of the designed structures is development of low-noise, low-threshold novel light sources and sensitive biosensors, as well as providing useful analogies for the microwave and acoustic devices and components.

ACKNOWLEDGEMENT

The author would like to thank Prof. A. I. Nosich and Miss E. Smotrova of the Institute of Radiophysics and Electronics NASU (Kharkov, Ukraine) for fruitful discussions.

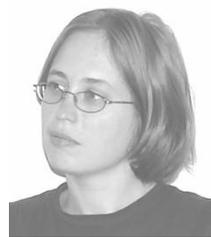

**Svetlana V. Boriskina** (S'96-M'01) was born in Kharkov, Ukraine in 1973. She received the M.Sc. degree with honours in radio physics and Ph.D. degree in physics and mathematics from Kharkov National University, Ukraine, in 1995 and 1999, respectively.

Since 1997 she has been a researcher in the School of Radiophysics at the V. Karazin Kharkov National University, and from 2000 to 2004, a Royal Society – NATO Postdoctoral Fellow and a Research Fellow in the School of Electrical and Electronic Engineering, University of Nottingham, UK. Her research interests are in integral equation methods for electromagnetic wave scattering and eigenvalue problems, with applications to optical fibers and waveguides, WG-mode microdisk lasers, and microring resonator filters.